\documentclass[cmp]{svjour}  
\usepackage{amsmath}
\usepackage{amsfonts,amssymb}

\journalname{Communications in Mathematical Physics}

\begin{document}

\newcommand{\nl}{\nonumber\\}
\newcommand{\nnl}{\nl[6mm]}
\newcommand{\nle}{\nl[-2.5mm]\\[-2.5mm]}
\newcommand{\nlb}[1]{\nl[-2.5mm]\label{#1}\\[-2.5mm]}
\newcommand{\bl}{&&\quad}
\newcommand{\ab}{\allowbreak}

\renewcommand{\leq}{\leqslant}
\renewcommand{\geq}{\geqslant}

\renewcommand{\theequation}{\thesection.\arabic{equation}}
\let\ssection=\section
\renewcommand{\section}{\setcounter{equation}{0}\ssection}

\newcommand{\be}{\begin{equation}}
\newcommand{\ee}{\end{equation}}
\newcommand{\bes}{\begin{eqnarray}}
\newcommand{\ees}{\end{eqnarray}}
\newcommand{\eens}{\nonumber\end{eqnarray}}

\renewcommand{\/}{\over}
\renewcommand{\d}{\partial}
\newcommand{\dd}{{\check\d}}
\newcommand{\dds}{{d\/ds}}
\newcommand{\ddt}{{d\/dt}}

\newcommand{\eps}{\epsilon}
\newcommand{\dlt}{\delta}
\newcommand{\al}{\alpha}
\newcommand{\bt}{\beta}
\newcommand{\si}{\sigma}
\newcommand{\la}{\lambda}
\newcommand{\ka}{\kappa}
\newcommand{\ups}{\upsilon}

\newcommand{\tdiff}{{\widetilde{diff}}}
\newcommand{\tmap}{{\widetilde{map}}}

\newcommand{\xmu}{\xi^\mu}
\newcommand{\xnu}{\xi^\nu}
\newcommand{\xz}{\xi^0}
\newcommand{\ynu}{\eta^\nu}
\newcommand{\yz}{\eta^0}
\newcommand{\dmu}{\d_\mu}
\newcommand{\dnu}{\d_\nu}
\newcommand{\dsi}{\d_\si}
\newcommand{\dotau}{\d_\tau}
\newcommand{\drho}{\d_\rho}
\newcommand{\qmu}{q^\mu}
\newcommand{\qnu}{q^\nu}
\newcommand{\qsi}{q^\si}
\newcommand{\qtau}{q^\tau}
\newcommand{\qrho}{q^\rho}
\newcommand{\pmu}{p_\mu}
\newcommand{\pnu}{p_\nu}

\newcommand{\phim}{\phi_{,\mm}}
\newcommand{\phin}{\phi_{,\nn}}
\newcommand{\pha}{\phi_\al}
\newcommand{\pham}{\phi_{\al,\mm}}
\newcommand{\pim}{\pi^{,\mm}}
\newcommand{\pin}{\pi^{,\nn}}
\newcommand{\pa}{\pi^\al}
\newcommand{\pam}{\pi^{\al,\mm}}
\newcommand{\hL}{{\hat L}}
\newcommand{\hXi}{{\hat \Xi}}

\newcommand{\tpi}{{1\/2\pi i}}
\newcommand{\fpi}{{1\/4\pi i}}
\newcommand{\half}{{1\/2}}
\newcommand{\Np}{{N+p\choose p}}
\newcommand{\Npone}{{N+p\choose p-1}}

\newcommand{\summp}{\sum_{|\mm|\leq p}}
\newcommand{\summn}{\sum_{|\mm|\leq|\nn|}}
\newcommand{\summnp}{\sum_{|\mm|\leq|\nn|\leq p}}
\newcommand{\summrn}{\sum_{|\mm|\leq|\rr|\leq|\nn|}}
\newcommand{\suml}{\sum_{\ell=0}^p}

\newcommand{\mm}{{\mathbf m}}
\newcommand{\nn}{{\mathbf n}}
\newcommand{\rr}{{\mathbf r}}
\newcommand{\noll}{{\mathbf 0}}
\newcommand{\mmu}{{\underline \mu}}
\newcommand{\nnu}{{\underline \nu}}

\newcommand{\virext}{{{c_4\/24\pi i}(\dddot\dlt(s-t) + \dot\dlt(s-t))}}

\renewcommand{\L}{{\cal L}}
\newcommand{\J}{{\cal J}}
\newcommand{\XX}{{\cal X}}
\newcommand{\YY}{{\cal Y}}
\newcommand{\MM}{{\cal M}}
\newcommand{\EE}{{\cal E}}
\newcommand{\FF}{{\cal F}}
\newcommand{\PP}{{\cal P}}
\newcommand{\QQ}{{\cal Q}}
\newcommand{\Lxi}{\L_\xi}
\newcommand{\Leta}{\L_\eta}
\newcommand{\tot}{{TOT}}
\newcommand{\gh}{{{\rm gh}}}

\newcommand{\into}{\hookrightarrow}
\newcommand{\e}{{\rm e}}
\newcommand{\id}{{\rm id}}
\newcommand{\arroww}[1]{{\ \stackrel{#1}{\longrightarrow}\ }}

\newcommand{\rep}{{\varrho}}

\newcommand{\trS}{{\rm tr}_{S_\ell}\kern0.13mm}
\newcommand{\tr}{{\rm tr}\kern0.7mm}
\newcommand{\dimm}{{\rm dim}\kern0.13mm}
\newcommand{\oj}{{\mathfrak g}}

\newcommand{\mum}{{\mu_1..\mu_m}}
\newcommand{\nun}{{\nu_1..\nu_n}}
\newcommand{\rnun}{{\rho|\nu_1..\nu_n}}
\newcommand{\nux}[1]{{\nu_1..#1..\nu_n}}
\newcommand{\nunj}{{\nu_1..\check\nu_j..\nu_n}}

\newcommand{\bxy}{{[\xi,\eta]}}
\newcommand{\bra}[1]{\big{\langle}#1\big{|}}
\newcommand{\ket}[1]{\big{|}#1\big{\rangle}}

\newcommand{\toplus}{{\widetilde{\oplus}}}
\newcommand{\tltimes}{{\widetilde{\ltimes}}}
\newcommand{\km}[1]{{\widehat{#1}}}
\newcommand{\no}[1]{{\,:\kern-0.7mm #1\kern-1.2mm:\,}}

\newcommand{\RR}{{\mathbb R}}
\newcommand{\CC}{{\mathbb C}}
\newcommand{\ZZ}{{\mathbb Z}}
\newcommand{\NN}{{\mathbb N}}

\title{{Extended Diffeomorphism Algebras and Trajectories In Jet Space}}
\titlerunning{Extended Diffeomorphism Algebras}

\author{T. A. Larsson}
\institute{Vanadisv\"agen 29, S-113 23 Stockholm, Sweden\\
 \email{tal@hdd.se}}
\authorrunning{T. A. Larsson}

\date{Received: 29 October 1998 / Accepted: 2 May 2000}
\communicated{T. Miwa}

\maketitle
\begin{abstract}
Let the DRO (Diffeomorphism, Reparametrization, Observer) algebra
$DRO(N)$ be the extension of $diff(N)\oplus diff(1)$ by its four
inequivalent Virasoro-like cocycles. Here
$diff(N)$ is the diffeomorphism algebra in $N$-dimen\-sional spacetime 
and $diff(1)$ describes reparametrizations of trajectories in 
the space of tensor-valued $p$-jets.
$DRO(N)$ has a Fock module for each $p$ and each representation of 
$gl(N)$. Analogous representations for gauge algebras (higher-dimensional
Kac-Moody algebras) are also given.
The reparametrization symmetry can be eliminated by a gauge fixing
procedure, resulting in previously discovered modules. In this
process, two $DRO(N)$ cocycles transmute into anisotropic cocycles
for $diff(N)$. Thus the Fock modules of toroidal Lie algebras and 
their derivation algebras are geometrically explained.
\end{abstract}

\section{Introduction}
Consider the algebra of diffeomorphisms in $N$-dimen\-sional spacetime, 
$diff(N)$. The classical representations act on tensor densities 
over spacetime \cite{ER96,Rud74}, but this is not a good starting point
for quantization. Na\"\i vely, one would try to introduce canonical
momenta and normal order, but this only works in one dimension, where
this procedure gives Fock representations of the Virasoro algebra.
In higher dimensions, infinities are encountered; formally, a central
extension proportional to the number of time-independent functions arises.
Moreover, $diff(N)$ has no central extension when $N>1$. 

$diff(N)$ acts naturally on the corresponding space of $p$-jets, $p$ 
finite.
The infinite jet space is essentially the space of functions, insofar
as functions may be identified with their Taylor series. 
This realization of $diff(N)$ is finite-dimen\-sional but non-linear;
diffeomorphisms act linearly on the Taylor coefficients with matrices 
depending non-linearly on the base point.
The corresponding Fock representation is well defined but not very 
interesting, because it gives us back the original tensor densities
(and derivatives thereof), and no extensions arise.

To remedy this, consider the space of trajectories in jet space.
$diff(N)$ acts naturally on this space as well, but in a highly reducible
fashion; the realization is a continuous direct sum because every point 
on a trajectory transforms independently of its neighbors. 
This degeneracy can be lifted by adding an extra $diff(1)$ factor 
describing reparametrizations, and thus the total algebra is 
$diff(N)\oplus diff(1)$. The DRO algebra $DRO(N)$ is the extension of 
this algebra by its four independent Virasoro-like cocycles,
which are non-central except in one dimension.
The canonical normal ordering with respect to reparametrizations
results in Fock modules for $DRO(N)$.
On the group level, this corresponds to a representation up to a 
local phase; only if the phase is globally constant, the Lie algebra
extension is central.

Reparametrizations are then eliminated by Hamiltonian reduction.
Since they generate first class constraints, a gauge fixing condition
must be introduced; a natural choice is to identify 
one coordinate with the parameter along the trajectory. 
Poisson brackets are now replaced with Dirac brackets before 
normal ordering. This yields a projective 
realization of $diff(N)$, which was discovered by hand in \cite{Lar97a} 
(that paper was limited to zero-jets). 
In particular, two of the $diff(N)\oplus diff(1)$ cocycles transmute
into the anisotropic $diff(N)$ extensions described in that paper.
By further specialization to scalar-valued jets (and choosing a 
Fourier basis on the $N$-dimen\-sional torus), we recover
the results of \cite{ERM94} on the derivation algebra of toroidal Lie
algebras.
I thus give a complete geometrical explanation of the rather surprising 
results in \cite{ERM94,Lar97a}, and generalize in two ways: 
reparametrizations are separated from diffeomorphisms, 
and arbitrary tensor-valued $p$-jets are considered, not only zero-jets. 

Berman and Billig \cite{BB98} independently studied tensor-valued
objects, but only as modules over the ``spatial'' subalgebra $diff(N-1)$.
For a supersymmetric generalization, see \cite{Lar97b}.
Proper representations were studied in \cite{ER96}.

It was noted by several authors \cite{BB98,Lar97b} that the gauge-fixed
algebra is ``space-time asymmetric'' in the sense that time is a
distinguished direction. In the present work this anisotropy is
isolated in the gauge fixing condition, whereas the underlying algebraic
structure is completely isotropic.

The gauge algebra $map(N,\oj)$, i.e. the algebra of maps from 
$N$-dimen\-sional spacetime to a finite-dimen\-sional Lie algebra $\oj$, 
has similar projective representations. 
This representation theory is also developed in 
the present paper, and thus the results of 
\cite{BB98,Bil97,EMY92,FM94,Lar97a,MEY90}
on toroidal Lie algebras are geometrically explained and generalized.

All considerations in this paper are local, but I expect that the results 
can be globalized without too much difficulty.
It is clear that the first de Rham homology plays an important r\^ole, 
both because the basic objects are one-dimen\-sional trajectories 
and because closed one-chains appear in (\ref{tdiff}) below.

\section{ The Algebra $DRO(N)$ }
\label{sec:DRO}
Let $\xi=\xmu(x)\dmu$, $x\in\RR^N$, $\dmu = \d/\d x^\mu$,
be a vector field, with commutator 
$[\xi,\eta] \equiv \xmu\dmu\ynu\dnu - \ynu\dnu\xmu\dmu$.
Greek indices $\mu,\nu = 0,1,..,N\!-\!1$ label the 
spacetime coordinates and the summation convention is used on all kinds 
of indices.
The diffeomorphism algebra (algebra of vector fields, Witt algebra) 
$diff(N)$ is generated by Lie derivatives $\Lxi$.
In particular, we refer to diffeomorphisms on the circle as 
repara\-metrizations. They form an additional $diff(1)$ algebra with 
generators $L(t)$, $t\in S^1$. $diff(N)\oplus diff(1)$ is the Lie
algebra with brackets
\bes
[\Lxi,\Leta] &=& \L_\bxy, \nl
{[}L(s),\Lxi] &=& 0, \label{diff}\\
{[}L(s), L(t)] &=& (L(s) + L(t)) \dot\dlt(s-t).
\eens
Alternatively, we describe reparametrizations in terms of generators
$L_f$, where $f = f(t)d/dt$ is a vector field on the circle:
\[
L_f = \int dt\ f(t) L(t).
\]
The commutator is $[f,g] = (f\dot g - g\dot f)d/dt$,
where a dot indicates the $t$ derivative.
The assumption that $t\in S^1$ is for technical simplicity; it enables
jets to be expanded in a Fourier series, but it is physically
quite unjustified because it means that spacetime is periodic in
the time direction. However, all we really need is that
$\int dt\ \dot F(t) = 0$ for all functions $F(t)$. Most results are
unchanged if we instead take $t\in\RR$ and replace Fourier sums with
Fourier integrals everywhere.

Introduce $N$ priviledged functions on the circle $\qmu(t)$,
which can be interpreted as the trajectory of an observer (or base point).
Let the observer algebra $Obs(N)$ be
the space of local functionals of $\qmu(t)$, i.e. polynomial functions of  
$\qmu(t)$, $\dot\qmu(t)$, ... $d^k \qmu(t)/dt^k$,  $k$ finite, 
regarded as a commutative Lie algebra.
The {\em DRO (Diffeomorphism, Repara\-metrization, Observer)} algebra 
$DRO(N)$ is an abelian but non-central Lie algebra extension of
$diff(N)\oplus diff(1)$ by $Obs(N)$:
\bes
0 \longrightarrow Obs(N) \longrightarrow DRO(N) \longrightarrow
 diff(N)\oplus diff(1) \longrightarrow 0.
\label{seq}
\ees
The extension depends on the four parameters $c_j$, $j = 1, 2, 3, 4$,
to be called  {\em abelian charges}; the name is chosen in analogy
with the central charge of the Virasoro algebra.
The sequence (\ref{seq}) splits ($DRO(N)$ is a semi-direct product)
iff all four abelian charges vanish. The brackets are given by
\bes
[\Lxi,\Leta] &=& \L_{[\xi,\eta]} 
 + \tpi\int dt\ \dot\qrho(t) 
 \Big( c_1 \drho\dnu\xmu(q(t))\dmu\ynu(q(t)) +\nl
 \bl+ c_2 \drho\dmu\xmu(q(t))\dnu\ynu(q(t)) \Big), \nl
{[}L_f, \Lxi] &=& {1\/4\pi i} \int dt\ 
 (c_3\ddot f(t) - ia_3\dot f(t))\dmu\xmu(q(t)), \nl
{[}L_f,L_g] &=& L_{[f,g]} 
 + {c_4\/24\pi i}\int dt (\ddot f(t) \dot g(t) - \dot f(t) g(t)), 
\label{DRO}\\
{[}\Lxi,\qmu(t)] &=& \xmu(q(t)), \nl
{[}L_f,\qmu(t)] &=& -f(t)\dot\qmu(t), \nl
{[}\qmu(s), \qnu(t)] &=& 0,
\eens
extended to all of $Obs(N)$ by Leibniz' rule and linearity.
The parameter $a_3$ is cohomologically trivial and can be removed by 
the redefinition
\bes
\Lxi \to \Lxi + {ia_3\/4\pi i} \int dt\ \dmu\xmu(q(t)).
\ees
The remaining four cocycles are non-trivial. 
We identify $c_4$ as the central 
charge in the Virasoro algebra generated by repara\-metrizations.

It is not difficult to reformulate the DRO algebra as a proper Lie 
algebra, by introducing a compete basis for $Obs(N)$. 
In fact, it suffices to consider the two linear operators
$S_0(F)$ and $S^\rho_1(F_\rho)$, defined for two arbitrary functions
$F(t,x)$ and $F_\rho(t,x)$, $t\in S^1$, $x\in\RR^N$,
\bes
S_0(F) &=& \tpi \int dt\ F(t,q(t)), \nle
S^\rho_1(F_\rho) &=& \tpi \int dt\ \dot\qrho(t) F_\rho(t,q(t)).
\eens
$DRO(N)$ now takes the form
\bes
&&[\Lxi,\Leta] = \L_\bxy + S_1^\rho(c_1 \drho\dnu\xmu\dmu\ynu 
 + c_2 \drho\dmu\xmu\dnu\ynu), \nl
&&{[}L_f,\Lxi] = {1\/2} S_0((c_3\ddot f - ia_3\dot f)\dmu\xmu), \nl
&&{[}L_f,L_g] = L_{f\dot g-\dot fg} + {c_4\/12}S_0(\ddot f\dot g-\dot fg), \nl
&&{[}\Lxi, S_0(F)] = S_0(\xmu\dmu F), \nl
&&{[}L_f, S_0(F)] = S_0(f{\d F\/\d t} + \dot f F), 
\label{tdiff}\\
&&{[}\Lxi, S^\rho_1(F_\rho)] = 
 S^\rho_1(\xmu\dmu F_\rho + \drho\xmu F_\mu), \nl
&&{[}L_f, S^\rho_1(F_\rho)] = S^\rho_1(f{\d F_\rho\/\d t}), \nl
&& S_0({\d F\/\d t}) + S_1^\mu(\dmu F) \equiv 0, \nl
&&S_0(f) = \tpi \int dt\ f(t), \qquad
\hbox{if $f(t)$ is independent of $x$,}
\eens
where $\dot f = df/dt$. That (\ref{tdiff}) defines a Lie algebra
follows from the explicit realization in Theorem \ref{Lthm} below,
but it is also straightforward to verify the Jacobi identities.

If $F_\rho(t,x)$ is independent of $t$, $S_1^\rho(F_\rho)$ is dual to
closed one-forms, and as such it can be viewed as a closed one-chain.
Dzhumadil'daev \cite{Dzhu96} has given a list of $diff(N)$
extensions by modules of tensor fields; see also \cite{Lar00}.
The cocycles $c_1$ and $c_2$ are related to his cocycles $\psi^W_4$ and 
$\psi^W_3$, respectively. In fact, they are equal for $S_1^\rho$ an
exact one-chain, but closed one-chains are not included in
Dzhumadil'daev's list since they are not tensor modules.
In one dimension,  $S_1(f) = 1/(2\pi i) \int dx^0\ f(x^0)$,
so the first line in (\ref{tdiff}) reduces to the Virasoro algebra with 
central charge $c = 12(c_1+c_2)$.

The cocycles in $DRO(N)$ have a natural origin from the diffeomorphism
algebra in $(N+1)$-dimensional space. Let its coordinates be
$z^A$, where capital indices $A= -1,0,1,2,...,N$ run over $N+1$ values
and the extra direction is labelled by $-1$. 
$diff(N+1)$ has an abelian extension with two Virasoro-like cocycles:
\bes
[\L_\XX, \L_\YY] &=& \L_{[\XX,\YY]} +
 S^C(c_1 \d_C\d_B\XX^A\d_A\YY^B + c_2 \d_C\d_A\XX^A\d_B\YY^B), \nl
{[}\L_\XX, S^C(F_C)] &=& S^C(\XX^A\d_AF_C + \d_C\XX^AF_A), 
\label{diff(N+1)}\\
S^C(\d_C F) &\equiv& 0,
\eens
where $\XX = \XX^A(z)\d_A$ is an $(N+1)$-dimensional vector field.
The cocycles multiplying $c_1$ and $c_2$ are simply those found by
Eswara Rao and Moody \cite{ERM94} and myself \cite{Lar91}, in one
extra dimension.
Now embed $diff(N)\oplus diff(1)\subset diff(N+1)$ in the natural
manner: $z^\mu=x^\mu, z^{-1} = t$, $\XX^A(z) = (\xmu(x), f(t))$,
$\L_\XX = (\Lxi, L_f)$, $S^C(F_C) = S_0(F) + S^\rho_1(F_\rho)$,
where $F=F_{-1}$.
Under this decomposition, (\ref{diff(N+1)}) restricts to (\ref{tdiff})
with $c_3 = 2c_2$, $c_4 = 12(c_1+c_2)$, up to a trivial cocycle.
However, it is easy to see that $c_3$ and $c_4$ are in fact independent
parameters, so there are four different cocycles in total.
In Sect. \ref{sec:constraints} below I will show that the complicated
anisotropic cocycles in \cite{Lar97a} can be obtained from (\ref{DRO})
by a gauge-fixing procedure.

We are interested in representations of $DRO(N)$ that are of 
lowest energy type with respect to the {\em Hamiltonian}
\bes
H = L_{-i\ddt} = -i\int dt\ L(t).
\label{Ham}
\ees
Such a representation contains a cyclic state $\ket0$ (the {\em vacuum}), 
satisfying 
\bes
H\ket0 = h\ket0, 
\label{le}
\ees
and $A\ket0 = 0$
for every operator $A$ such that $[H,A] = -w_AA$, $w_A>0$.
The lowest energy $h$ also characterizes the representation.

\section{ Preliminaries }
Consider the space of $V$-valued functions over spacetime, where $V$
carries an $gl(N)$ representation $\rep$. This is our
configuration space which will be denoted by $\QQ$.
A basis is given by $\pha(x)$, $x\in\RR^N$, 
where the index $\al$ labels different components of tensor densities.
The fields can be either bosonic or fermionic, but it is assumed that
all components have the same parity.
Let $\mm = (m_0, \ab m_1, \ab .., \ab m_{N-1})$, all $m_\mu\geq0$, be a 
multi-index of length
$|\mm| = \sum_{\mu=0}^{N-1} m_\mu$, let $\mmu$ be a unit vector in the 
$\mu^{{\rm th}}$ direction, and let $0$ be the multi-index of length 
zero. Denote by  
\bes
\d_\mm\pha(x) = \underbrace{\d_0 .. \d_0}_{m_0} .. 
 \underbrace{\d_{N-1} .. \d_{N-1}}_{m_{N-1}} \pha(x).
\label{dmm}
\ees
Diffeomorphisms act as follows on derivatives of tensor densities:
\bes
[\Lxi,\d_\nn\phi(x)] &=& 
\d_\nn( -\xmu(x)\dmu\phi(x) - \dnu\xmu\rep(T^\nu_\mu)\phi(x) )
\nlb{tensor}
&=& - \xmu(x)\d_{\nn+\mmu}\phi(x)
- \summn T^\mm_\nn(\xi(x)) \d_\mm\phi(x), \nl
T^\mm_\nn(\xi) &=& 
 {\nn\choose\mm} \d_{\nn-\mm+\nnu}\xmu\rep(T^\nu_\mu) \nl
 &+& {\nn\choose\mm-\mmu} (1-\dlt^{\mm-\mmu}_\nn)
 \d_{\nn-\mm+\mmu}\xmu \rep(1),
\label{Tmn}
\ees
where the $V$ index ($\al$) was suppressed,
$\mm! = m_0! m_1! .. m_{N-1}!$ and 
\break ${\nn\choose\mm} = \ab \nn!/\ab\mm!\ab(\nn-\mm)!$.
Our convention is that $gl(N)$ has basis $T^\mu_\nu$ and brackets
\bes
[T^\mu_\nu,T^\si_\tau] = 
\dlt^\si_\nu T^\mu_\tau - \dlt^\mu_\tau T^\si_\mu.
\label{glN}
\ees
$\rep(T^\nu_\mu)$ are the matrices in the representation $\rep$,
acting on a tensor density with $p$ upper and $q$ lower indices and
weight $\ka$ as follows:
\bes
\rep(T^\mu_\nu)\phi^{\si_1..\si_p}_{\tau_1..\tau_q}
= -\ka \dlt^\mu_\nu \phi^{\si_1..\si_p}_{\tau_1..\tau_q} 
+\sum_{i=1}^p \dlt^{\si_i}_\nu 
 \phi^{\si_1..\mu..\si_p}_{\tau_1..\tau_q}
-\sum_{j=1}^q \dlt^\mu_{\tau_j}
 \phi^{\si_1..\si_p}_{\tau_1..\nu..\tau_q}.
\label{gltens}
\ees
The matrices $T^\mm_\nn(\xi)$, with components $T^\mm_\nn(\xi)^\al_\bt$,
satisfy
\bes
T^\mm_{\nn+\nnu}(\xi) &=& \dnu\xmu\dlt^\mm_{\nn+\mmu} 
 + T^\mm_\nn(\dnu\xi) + T^{\mm-\nnu}_\nn(\xi), \nl
T^\mm_0(\xi) &=& \dlt^\mm_0 \dnu\xmu \rep(T^\nu_\mu), \nl
\dnu T^\mm_\nn(\xi) &=& T^\mm_\nn(\dnu\xi),
\label{td} \\
T^\mm_\nn([\xi,\eta]) &=& 
\xmu T^\mm_\nn(\dmu\eta) - \ynu T^\mm_\nn(\dnu\xi) \nl
&& + \summrn (T^\rr_\nn(\xi) T^\mm_\rr(\eta) 
- T^\rr_\nn(\eta) T^\mm_\rr(\xi) ).
\eens
In particular, $T^\mm_\nn(\xi)=0$ if $|\mm|>|\nn|$.

Let $\tr$ denote the trace in $gl(N)$ representation $\rep$.
Define numbers $\dimm(\rep)$, $k_0(\rep)$, $k_1(\rep)$, $k_2(\rep)$ by
\bes
\tr1 &=& \dimm(\rep), \nl
\tr T^\mu_\nu &=& k_0(\rep) \dlt^\mu_\nu, 
\label{kdef}\\
\tr T^\mu_\nu T^\si_\tau &=&
 k_1(\rep) \dlt^\mu_\tau \dlt^\si_\nu 
 + k_2(\rep) \dlt^\mu_\nu \dlt^\si_\tau.
\eens
For an unconstrained tensor transforming as in (\ref{gltens}),
\bes
\dimm(\rep) = N^{p+q}, &\qquad&
k_0(\rep)= -(p-q-\ka N) N^{p+q-1},
\label{krep}\\
k_1(\rep) = (p+q)N^{p+q-1}, &\qquad&
k_2(\rep) = ((p-q-\ka N)^2 - p - q) N^{p+q-2}.
\eens
Note that if $\ka = (p-q)/N$, $\rep$ is an $sl(N)$ representation.
Let $S_\ell$ be the symmetric representation on $\ell$ lower indices,
appropriate for multi-indices. 
We have $\dimm(S_\ell) = \sum_\mm \dlt^\mm_\mm$, etc.,
where
\bes
\dimm(S_\ell) = {N-1+\ell\choose\ell}, &\qquad&
k_0(S_\ell) = {N-1+\ell\choose\ell-1},
\nlb{kSl}
k_1(S_\ell)= {N+\ell\choose\ell-1}, &\qquad&
k_2(S_\ell) = {N-1+\ell\choose\ell-2}.
\eens

\begin{lemma}\label{Tlemma}
\bes
&i.& \summp \dlt^\mm_\mm\, \tr 1 = \Np \dimm(\rep), \nl
&ii.&\summp \tr T^\mm_\mm(\xi)
= \dmu\xmu (\Np k_0(\rep) + \Npone \dimm(\rep) ), \nl
&iii.&\sum_{\scriptstyle|\mm|\leq|\nn|\leq p 
  \atop \scriptstyle|\nn|\leq|\mm|\leq p}
 \tr T^\mm_\nn(\xi) T^\nn_\mm(\eta)
= \dnu\xmu\dmu\ynu(\Np k_1(\rep) + {N+p+1\choose p-1} \dimm(\rep) ) \nl
&&+ \dmu\xmu\dnu\ynu( \Np k_2(\rep) +{N+p\choose p-2}\dimm(\rep) 
 + 2\Npone k_0(\rep) ).
\eens
\end{lemma}

\begin{proof} 
If $|\mm|=|\nn|=\ell$,
\bes
T^\mm_\nn(\xi) = \dmu\xi^\nu(\rep(T^\mu_\nu)\dlt^\mm_\nn 
 + \rep(1)\zeta^\mm_\nn(T^\mu_\nu)),
\label{Ttop}
\ees
where $\zeta^\mm_\nn(T^\mu_\nu)$ are the representation matrices in
$S_\ell$, acting on multi-indices. Only the top values (\ref{Ttop}) 
contribute to the traces, which means 
that we can ignore that higher jets do not transform as 
$S_\ell$-valued zero-jets. By the definition (\ref{kdef}) and (\ref{Ttop}),
\bes
\dimm(\rep\otimes S_\ell) &=& \dimm(\rep)\cdot\dimm(S_\ell), \nl
k_0(\rep\otimes S_\ell) &=& 
k_0(\rep)\dimm(S_\ell) + \dimm(\rep)\,k_0(S_\ell), \nle
k_1(\rep\otimes S_\ell) &=& 
k_1(\rep)\dimm(S_\ell) + \dimm(\rep)\,k_1(S_\ell), \nl
k_2(\rep\otimes S_\ell) &=& 
k_2(\rep)\dimm(S_\ell) + \dimm(\rep)\,k_2(S_\ell)
+ 2k_0(\rep)k_0(S_\ell).
\eens
The lemma now follows from (\ref{kSl}) and the following sums:
\bes
\suml \dimm(S_\ell) = \Np, &\qquad&
\suml k_0(S_\ell) = \Npone,
\nlb{kSp}
\suml k_1(S_\ell) = {N+p+1\choose p-1}, &\qquad&
\suml k_2(S_\ell) = {N+p\choose p-2}.
\eens
\qed
\end{proof}

\section{Jet Space Trajectories}
Let $J^p\QQ$ be the space of trajectories in the space of
$V$-valued $p$-jets, 
with coordinates $(\qmu(t), \pham(t))$, where $|\mm| \leq p$ 
and $t\in S^1$. The parameter $t$ is referred to as {\em time} and
$\qmu(t)$ as the observer's trajectory in spacetime.
$DRO(N)$ acts on $J^p\QQ$ as follows:
\bes
[\Lxi,\phin(t)] &=& 
 -\summn T^\mm_\nn(\xi(q(t))) \phim(t), \nl
{[}L(s),\phin(t)] &=& -\dot\phin(t) \dlt(s-t) 
 + \la\phin(t)\dot\dlt(s-t) + iw\phin(t)\dlt(s-t),
\nlb{Lphi}
{[}\Lxi,\qmu(t)] &=& \xmu(q(t)), \nl
{[}L(s),\qmu(t)] &=& -\dot\qmu(t)\dlt(s-t).
\eens

Clearly, there is a chain of inclusions 
$J^{-1}\QQ\subset J^0\QQ \subset J^1\QQ \subset \ldots$,
where $J^{-1}\QQ$ consists of $\qmu(t)$ only.
Hence $J^p\QQ$ is reducible (but indecomposable) as a $DRO(N)$ 
realization.
This kind of reducibility is not present in the Fock modules below, 
because jets of all orders up to $p$ are
created from the vacuum, cf. (\ref{create}).

We call $\la$ the {\em causal weight} of $\phi$, 
in contradistinction to its tensorial weight $\ka$.
The {\em shift parameter} $w$ can be eliminated by the redefinition
\bes
\phin(t) \to \e^{-iwt} \phin(t),
\ees
so it is only defined up to an integer. 
The triple $(\ka,\la,w)$ will collectively be referred
to as the {\em weights} of $\phi$. The observer's trajectory $\qmu(t)$ 
has causal weight $0$ but it does not transform 
as a zero-jet under diffeomorphisms. However, its time derivative
has causal weight $1$ and {\em does} transform as a 
(vector-valued) zero-jet,
\bes
[\Lxi,\dot\qmu(t)] &=& \dnu\xmu(q(t))\dot\qnu(t),
\nlb{Lqt}
{[}L(s),\dot\qmu(t)] &=& -\ddot \qmu(t)\dlt(s-t) + \dot\qmu(t)\dot\dlt(s-t).
\eens

A point in $J^\infty\QQ$ can be identified with a trajectory in the 
space of $V$-valued functions via generating functions; 
for $x = (x^\mu)\in\RR^N$, define
\bes
\pha(x,t) = \sum_{|\mm|\geq0} {1\/\mm!} 
 \pham(t)(x-q(t))^\mm,
\label{gen}
\ees
where
\bes
(x-q(t))^\mm = (x^0-q^0(t))^{m_0} (x^1-q^1(t))^{m_1} ..
 (x^{N-1}-q^{N-1}(t))^{m_{N-1}}.
\ees
$\pha(x,t)$ transforms as in (\ref{tensor}) under diffeomorphisms 
and as (\ref{Lphi}) under repara\-metrizations; note that
\bes
\ddt\pha(x,t) = \sum_{|\mm|\geq0} {1\/\mm!} 
(\dot\pham(t)(x-q(t))^\mm 
- m_\mu\dot\qmu(t)\pham(t)(x-q(t))^{\mm-\mmu}.
\eens
Moreover, $\d_\mm \pha(x,t) = \pham(x,t)$.
This formula suggests that we define a map
\bes
\dd_\mu: && J^p\QQ \longrightarrow J^{p+1}\QQ, \nl
\dd_\mu \qnu(t) &=& \dlt^\nu_\mu,
\label{dm}\\
\dd_\mu \phin(t) &=& \phi_{,\nn+\mmu}(t),
\eens
extended to the whole of $J^p\QQ$ by Leibniz' rule and linearity.
Further, define $\dd_\mm$ as in (\ref{dmm}). 
This operator satisfies
$\dd_\mm f(q(t)) = \d_\mm f(q(t))$ and
\bes
\dd_\mu\Lxi &=& \Lxi\dd_\mu + \dmu\xnu\dd_\nu,
\nle
\dd_\mu L(s) &=& L(s)\dd_\mu,
\eens
when acting on arbitrary functions on $J^p\QQ$.

\section{Realization in Fock Space}

Consider the symplectic space $J^p\PP$ obtained by adjoining 
to $J^p\QQ$ dual coordinates (jet momenta) $(\pmu(t), \pam(t))$.
The graded Poisson algebra $C^\infty(J^p\PP)$ is the associative, 
graded commutative algebra on symbols 
$(\qmu(t), \pham(t), \pmu(t),\ab \pam(t))$,
equipped with a compatible graded Lie structure: the Poisson bracket. 
The only non-zero brackets are
\bes
[\pmu(s), \qnu(t)] &=& \dlt^\nu_\mu \dlt(s-t), 
\label{Poisson} \\
{[}\pam(s), \phi_{\bt,\nn}(t)] &\equiv&
\mp {[}\phi_{\bt,\nn}(t), \pam(s)] 
 = \dlt^\mm_\nn \dlt_\bt^\al \dlt(s-t),
\eens
where we here and henceforth use the convention that the upper sign 
refers to bosons and the lower to fermions.

All functions over $S^1$ can be expanded in a Fourier series; e.g.
\bes
\pham(t) = \sum_{n=-\infty}^\infty
\hat\pham(n) \e^{-int} \equiv
\pham^<(t) + \hat\pham(0) + \pham^>(t),
\label{Fourier}
\ees
where $\pham^<(t)$ ($\pham^>(t)$) is the sum over 
negative (positive) frequency modes only.
$\hat\pham(0)$ will be referred to as the {\em zero mode}.
Quantization amounts to replacing the Poisson brackets (\ref{Poisson})
by graded commutators;
the Fock space $J^p\FF$ is the universal enveloping
algebra modulo relations
\bes
\qmu_<(t)\ket0 = \pmu^\leq(t)\ket0 = \pam_\leq(t)\ket0 
 = \pham^<(t)\ket0 = 0,
\ees
where
$\pmu^\leq(t) = \pmu^<(t)+ \hat\pmu(0)$ and
$\pam_\leq(t) = \pam_<(t) + \hat\pam(0)$.

Normal ordering is necessary to remove infinites and to obtain a well 
defined action on Fock space. Let $f(q(t),\phi(t))$ be a function of
$\qmu(t)$, $\phi(t)$, as well as its derivatives $\phim(t)$,
but independent of the canonical momenta.  
Denote 
\bes
\no{f(q(t),\phi(t))\pmu(t)} 
 &=& f(q(t),\phi(t))\pmu^\leq(t) + \pmu^>(t)f(q(t),\phi(t)), \\
\no{f_\bt^\al(q(t)) \pi^{\bt,\nn}(t) \pham(t) } &=&
\pi^{\bt,\nn}_>(t) f_\bt^\al(q(t)) \pham(t) 
\pm\pham(t) f_\bt^\al(q(t)) \pi^{\bt,\nn}_\leq(t).
\eens
In particular,
\bes
\no{ \pin(t) T^\mm_\nn(\xi) \phim(t) } 
= T^\mm_\nn(\xi)_\bt^\al  (\pi^{\bt,\nn}_>(t) \pham(t)
  \pm \pham(t) \pi^{\bt,\nn}_\leq(t) ).
\nonumber
\ees

We are now ready to state the main result.
\begin{theorem}
\label{Lthm}
The following operators provide a realization of 
$DRO(N)$ on $J^p\FF$:
\bes
\Lxi &=& \int dt\ \no{\xmu(q(t)) \pmu(t)}+ T(\xi(q(t)),t), \nl
T(\xi,t) &=& \mp \summnp 
 \no{ \pin(t) T^\mm_\nn(\xi) \phim(t) }, \nl
L(t) &=& - \no{\dot\qmu(t)\pmu(t)}+ L'(t), 
\label{emb} \\
L'(t) &=& \pm \summp \Big\{  - \no{ \pim(t)\dot\phim(t)} 
+ \la\no{ \ddt( \pim(t) \phim(t)) } \nl
&&+iw \no{ \pim(t) \phim(t)} \Big\}
\pm \Np \dimm(\rep){\la-\la^2-w+w^2\/4\pi i} , \nl
\eens
where the upper sign holds for bosons and the lower sign for fermions.
The abelian charges are
$c_3 = 1+c'_3$, $a_3 = 1+a'_3$, $c_1 = 1+c'_1$, $c_4=2N+c'_4$, where
\bes
c'_1 &=& \pm (\Np k_1(\rep) + {N+p+1\choose p-1} \dimm(\rep) ), \nl
c_2&=& \pm (\Np k_2(\rep) +{N+p\choose p-2} \dimm(\rep)
 + 2\Npone k_0(\rep) ), \nl
c'_3 &=& \pm (2\la-1)(\Np k_0(\rep) + \Npone \dimm(\rep) ), 
\label{cs} \\
a'_3 &=& \pm (2w-1)(\Np k_0(\rep) + \Npone \dimm(\rep) ), \nl
c'_4&=& \pm 2(1-6\la+6\la^2) \Np \dimm(\rep). 
\eens
$\dimm(\rep)$, $k_0(\rep)$, $k_1(\rep)$ and $k_2(\rep)$ were 
defined in (\ref{kdef}) and $\la$ and $w$ in (\ref{Lphi}). 
\end{theorem}

{F}rom (\ref{emb}) we read off the transformation laws for the jet momenta.
\bes
[\Lxi,\pnu(t)] &=& -\dnu\xmu(q(t)) \pmu(t) - T(\dnu\xi(q(t)),t), \nl
{[}L(s),\pnu(t)] &=& \pnu(s)\dot\dlt(s-t), \nl
{[}\Lxi,\pim(t)] &=& \summnp
 \pin(t) T^\mm_\nn(\xi(q(t))), 
\label{Lxp} \\
{[}L(s),\pim(t)] &=& -\dot\pim(t) \dlt(s-t) 
 + (1-\la)\pim(t)\dot\dlt(s-t) \nl
&&- iw\pim(t)\dlt(s-t).
\eens
Note the range of the sum, which depends on the
order of the jet. In particular, the top momentum $\pim(t)$,
$|\mm|=p$, transforms as a tensor-valued zero-jet.

Without normal ordering, Theorem \ref{Lthm} defines a proper but highly
reducible representation of $diff(N)$; in fact, it is a continuous 
direct sum of $p$-jets, one for each value of the time parameter $t$. 
This degeneracy is lifted by the introduction of the
reparametrization algebra. 
Using (\ref{dm}), the $diff(N)$ generators can be written as
\bes
\Lxi &=& \int dt\ \xmu(q(t)) 
 (\pmu(t) \pm \summp \pim(t)\phi_{,\mm+\mmu}(t) ) \nl
&&\mp \summp \pim(t) \dd_\mm(
 \xmu(q(t))\phi_{,\mmu} + \dnu\xmu(q(t))\rep(T^\nu_\mu) \phi(t) ).
\label{Lxid}
\ees
All formulas simplify for zero-jets. $T(\xi,t) =  \dnu \xmu T^\nu_\mu(t)$,
where $T^\nu_\mu(t)$ generate the Kac-Moody algebra $\km{gl(N)}$,
\bes
[T^\mu_\nu(s), T^\si_\tau(t)] &=& ( \dlt^\si_\nu T^\mu_\tau(s) 
 - \dlt^\mu_\tau T^\si_\nu(s) )\dlt(s-t) \nl
&&\mp \tpi ( k_1(\rep) \dlt^\mu_\tau \dlt^\si_\nu
 + k_2(\rep) \dlt^\mu_\nu \dlt^\si_\tau ) \dot\dlt(s-t), \\
{[}L'(s), T^\mu_\nu(t)] &=& T^\mu_\nu(s) \dot\dlt(s-t) 
 \mp {k_0(\rep)\/4\pi i}\dlt^\mu_\nu(\ddot\dlt(s-t) + i\dot\dlt(s-t)).
\eens

It should be stressed that the action in Theorem \ref{Lthm} on 
$J^p\FF$ is manifestly well defined, at least for the subalgebra of
polynomial vector fields. Namely, a monomial basis for $J^p\FF$ 
is given by 
finite strings in the non-negative modes $\hat \qmu(n)$, $\hat\pmu(n)$, 
$\hat\pham(n)$, $\hat\pam(n)$, $n \geq 0$, 
$|\mm|\leq p$, and a generic element is a finite linear combination 
of such monomials. For $\xi$ a polynomial vector field, finiteness is
preserved by (\ref{emb}).

Split the delta function into positive and negative frequency parts:
\bes
\dlt^>(t) = {1\/2\pi} \sum_{m>0} \e^{-imt}, \qquad
\dlt^\leq (t) = {1\/2\pi} \sum_{m\leq0} \e^{-imt}.
\ees
\begin{lemma}{\label{delta}} \cite{Lar97a}
\bes
&i.& \dlt^>(t) \dlt^\leq (-t) - \dlt^>(-t) \dlt^\leq (t)
= -\tpi\dot \dlt(t), \nl
&ii.& \dlt^>(t) \dot\dlt^\leq (-t) - \dot\dlt^>(-t) \dlt^\leq (t)
= \fpi (\ddot \dlt(t) + i\dot \dlt(t)), \nl
&iii.& \dot\dlt^>(t) \dot\dlt^\leq (-t) - \dot\dlt^>(-t)\dot\dlt^\leq(t)
= {1\/ 12\pi i} (\dddot \dlt(t) + \dot \dlt(t)). 
\eens
\end{lemma}

\begin{lemma}{\label{PPlemma}}
Let $\pi^A(s)$, $\phi_B(t)$, $s,t\in S^1$, 
generate a graded Heisenberg algebra, with non-zero brackets
$[\pi^A(s),\phi_B(t)] = \dlt^A_B\dlt(s-t)$. Then
\bes
[\pi^A_>(s),\phi_B(t)] = \dlt^A_B\dlt^>(s-t), &\qquad&
[\phi_B(s),\pi^A_>(t)] = \mp\dlt^A_B\dlt^>(t-s), \nl
{[}\pi^A_\leq(s),\phi_B(t)] = \dlt^A_B\dlt^\leq(s-t), &\qquad&
[\phi_B(s),\pi^A_\leq(t)] = \mp\dlt^A_B\dlt^\leq(t-s).
\eens
\end{lemma}

\begin{lemma}{\label{FElemma}}
Define
\bes
F(t) = \mp \no{ \pi^A(t) \dot\phi_A(t) }, \qquad
E^A_B(t) = \mp \no{ \pi^A(t) \phi_B(t) },
\label{FEdef}
\ees
where $\pi^A(s)$ and $\phi_B(t)$, defined as in the previous lemma,
carry the same statistics (so $E^A_B(t)$ is bosonic).
Then
\bes
[F(s),F(t)] &=& (F(s)+F(t))\dot\dlt(s-t) \nl 
&&\pm \dlt^A_A{1\/12\pi i}(\dddot\dlt(s-t)+\dot\dlt(s-t)),
\label{FF}\\
{[}F(s),E^A_B(t)] &=& E^A_B(s) \dot\dlt(s-t) 
\mp \dlt^A_B \fpi( \ddot\dlt(s-t) + i\dot\dlt(s-t)),
\label{FE}\\
{[}E^A_B(s),E^C_D(t)] &=& (\dlt^C_B E^A_D(s) 
 - \dlt^A_D E^C_B(s))\dlt(s-t) \nl
&&\mp\dlt^A_D \dlt^C_B \tpi\dot\dlt(s-t).
\label{EE}
\ees
\end{lemma}

\begin{proof} 
This lemma follows by direct calculation. 
The technique is illustrated for (\ref{FE}) only,
\bes
&&[F(s),E^A_B(t)] =
[\mp \pi^C_>(s)\dot\phi_C(s) - \dot\phi_C(s)\pi^C_\leq(s),
 \mp \pi^A_>(t)\phi_B(t) - \phi_B(t)\pi^A_\leq(t)] \nl
&&= \Big\{ \pi^A_>(s)\dds(\mp\dlt^>(t-s))\phi_B(t)
 \pm \pi^A_>(t)\dlt^>(s-t)\dot\phi_B(s) \Big\} \nl
\bl \pm\Big\{ \dlt^>(s-t)\pi^A_\leq(t)\dot\phi_B(s)
 \pm \pi^A_>(s)\phi_B(t) \dds(\mp\dlt^\leq(t-s)) \Big\} \nl
\bl \pm\Big\{ \pm\dot\phi_B(s)\pi^A_>(t)\dlt^\leq(s-t)
 + \dds(\mp\dlt^>(t-s))\phi_B(t)\pi^A_\leq(s) \Big\} \nl
\bl +\Big\{ \dot\phi_B(s)\dlt^\leq(s-t)\pi^A_\leq(t)
 \pm \phi_B(t)\dds(\mp\dlt^\leq(t-s))\pi^A_\leq(s) \Big\} \nl
&&= \mp\pi^A_>(s)\phi_B(t)\dds\dlt(t-s)
 \pm \pi^A_>(t)\dot\phi_B(s)\dlt(s-t) \\
\bl+\dds(\mp\dlt^>(t-s))\dlt^A_B\dlt^\leq(s-t) 
+ \dot\phi_B(s)\pi^A_\leq(t)\dlt(s-t) \nl
\bl\pm\dlt^>(s-t)\dds(\dlt^\leq(t-s))\dlt^A_B
 - \phi_B(t)\pi^A_\leq(s)\dds\dlt(t-s) \nl
&&= \mp\no{\pi^A(s)\phi_B(t)} \dds\dlt(t-s)
 \pm\no{\pi^A(t)\dot\phi_B(s)}\ \dlt(s-t) \nl
\bl \mp \dlt^A_B(\dlt^>(s-t)\dot\dlt^\leq(t-s)
 -\dot\dlt^>(t-s)\dlt^\leq(s-t)).
\eens
The result now follows by collecting terms and applying Lemma \ref{delta}
to obtain the central extension. 
\qed
\end{proof}

{\em Proof of Theorem \ref{Lthm}.}
First we note that in proving the brackets with $\qmu(t)$,
normal ordering is irrelevant because
$\Lxi$ is linear in $\pmu(t)$. This part is straightforward and not
given here.

We now turn to diffeomorphisms, and
set $\Lxi^0 = \int ds\ \no{\xmu(s)\pmu(s)}$, where we abbreviate
$\xmu(q(s)) = \xmu(s)$, etc.,
\bes
&&[\Lxi^0, \Leta^0] = \iint dsdt\ 
[\xmu(s)\pmu^\leq(s) + \pmu^>(s)\xmu(s),
 \ynu(t)\pnu^\leq(t) + \pnu^>(t)\ynu(t)] \nl
&&= \iint dsdt\
  \xmu(s)(\dmu\ynu(t)\dlt^\leq(s-t))\pnu^\leq(t)
 +\ynu(t)(-\dnu\xmu(s)\dlt^\leq(t-s))\pmu^\leq(s) \nl
\bl+\xmu(s)\pnu^>(t)(\dmu\ynu(t)\dlt^\leq(s-t))
 +(-\dnu\xmu(s)\dlt^>(t-s))\ynu(t)\pmu^\leq(s) \nl
\bl+(\dmu\ynu(t)\dlt^>(s-t))\pnu^\leq(t)\xmu(s)
 +\pmu^>(s)\ynu(t)(-\dnu\xmu(s)\dlt^\leq(t-s)) \nl
\bl+\pnu^>(t)(\dmu\ynu(t)\dlt^>(s-t))\xmu(s)
 +\pmu^>(s)(-\dnu\xmu(s)\dlt^>(t-s))\ynu(t) \nl
&&= \iint dsdt\ 
 \xmu(s)\dmu\ynu(t)\pnu^\leq(t)\dlt(s-t)
 + \dmu\ynu(t)\dlt^>(s-t)\dnu\xmu(s)\dlt^\leq(t-s) \nl
\bl+ \pmu^>(t)\xmu(s)\dmu\ynu(t)\dlt(s-t) 
 - \dnu\xmu(s)\dlt^>(t-s)\dmu\ynu(t)\dlt^\leq(s-t) \nl
\bl- \ynu(t)\dnu\xmu(s)\pmu^\leq(s)\dlt(s-t)
- \pmu^>(s)\ynu(t)\dnu\xmu(s)\dlt(s-t) \\
&&= \iint dsdt\ \no{ \xmu(s)\dmu\ynu(t)\pnu(t) }\ \dlt(s-t)
 - \no{ \ynu(t)\dnu\xmu(s)\pmu(s) } \dlt(t-s)\nl
\bl+ \dmu\ynu(t)\dnu\xmu(s)
 ( \dlt^>(s-t)\dlt^\leq(t-s) -  \dlt^>(t-s)\dlt^\leq(s-t) ).
\eens
We now apply Lemma \ref{delta} and integrate by parts, which yields
\bes
[\Lxi^0, \Leta^0] 
&=& \L^0_\bxy + \tpi\int ds\ \dnu\dot\xmu(s)\dmu\ynu(s) \nl
&=& \L^0_\bxy + S_1^\rho(\drho\dnu\xmu\dmu\ynu).
\ees
Let capital indices run over both tensor indices and multi-indices, 
e.g. $A=(\al,\mm)$, $\pi^A(s) = \pam(s)$,
$\phi_B(t) = \phi_{\bt,\mm}(t)$.
Now, $\Lxi = \Lxi^0 + \int dt\ T(\xi(q(t)),t)$, where
$T(\xi,t) = T^B_A(\xi) E^A_B(t)$ in an obvious notation.
$E^A_B(t)$ is defined in (\ref{FEdef}), the matrices $T^A_B(\xi)$
satisfy the relations (\ref{td}) and Lemma \ref{Tlemma}, with
\bes
\dlt^A_A &=& \summp \dlt^\mm_\mm, \nl
T^A_A(\xi) &=& \summp \tr T^\mm_\mm(\xi), \\
T^A_B(\xi)T^B_A(\eta) &=&
\sum_{\scriptstyle|\mm|\leq|\nn|\leq p 
 \atop \scriptstyle|\nn|\leq|\mm|\leq p}
 \tr T^\mm_\nn(\xi) T^\nn_\mm(\eta).
\eens
It follows from (\ref{EE}) that
\bes
&&{[}T(\xi(s),s),T(\eta(t),t)] \nl
&&=(T^C_A(\xi(s))T^B_C(\eta(t)) - T^B_C(\xi(s))T^C_A(\eta(t))) 
 E^A_B(s)\dlt(s-t) \nl
\bl\mp\tpi T^A_B(\xi(s)) T^B_A(\eta(t))\dot\dlt(s-t) \\
&&=( T([\xi(s),\eta(t)],s) - \xmu(s) T(\dmu\eta(t),s)
 + \ynu(t) T(\dnu\xi(s),s))\dlt(s-t)\nl
\bl-\tpi(c'_1\dnu\xmu(s)\dmu\ynu(t) 
 + c_2\dmu\xmu(s)\dnu\ynu(t))\dot\dlt(s-t),
\eens
where the parameters $c'_1$ and $c_2$ can now be computed
from Lemma \ref{Tlemma}, with the result (\ref{cs}).
Further,
\[
[\Lxi^0, T(\eta(t),t)] = \xmu(t) T(\dmu\eta(t),t),
\]
without extension. This concludes the proof for the $diff(N)$ subalgebra.

Next we turn to reparametrizations. They generate a Virasoro algebra
with central charge $c$, which may be written as
\bes
[L(s),L(t)] = (L(s)+L(t)) \dot\dlt(s-t) 
 + {c\/24\pi i}(\dddot\dlt(s-t) + \dot\dlt(s-t)).
\label{Vir}
\ees
Set $L^0(s) = -\no{ \dot\qmu(s)\pmu(s)}$.
This is recognized as being of the same form as $F(s)$ in 
Lemma \ref{FElemma}, with $N$ bosonic fields $\qmu(s)$, and thus they
generate a Virasoro algebra with central charge $2N$. Set
$L(t) = L^0(t) + L'(t)$, where
\bes
L'(t) = F(t) - \la \dot E^A_A(t) - iwE^A_A(t) 
 \pm \dlt^A_A {\la-\la^2-w+w^2\/4\pi i}.
\ees
By Lemma \ref{FElemma}, these operators generate a Virasoro algebra with
central charge $\pm 2(1-6\la+6\la^2) \dlt^A_A$.
Moreover, $[L^0(s),L'(t)] = 0$ and the parameter $c'$ in (\ref{cs})
follows from Lemma \ref{Tlemma}.

Finally, we want to prove that
\bes
[L(s),\Lxi] = \fpi(c_3\dmu\ddot\xmu(s) + ia_3\dmu\dot\xmu(s)).
\label{LsLx}
\ees
\bes
&&[L^0(s), \Lxi^0] = -\int dt\
[\dot\qmu(s)\pmu^\leq(s) + \pmu^>(s)\dot\qmu(s),
\xnu(t)\pnu^\leq(t) + \pnu^>(t)\xnu(t)] \nl
&&= -\int dt\ \dot\qmu(s)(\dmu\xnu(t)\dlt^\leq(s-t))\pnu^\leq(t)
- \xnu(t) \dds(\dlt^\mu_\nu\dlt^\leq(t-s))\pmu^\leq(s) \nl
\bl+ \dot\qmu(s)\pnu^>(t)(\dmu\xnu(t)\dlt^\leq(s-t))
- \dds(\dlt^\mu_\nu\dlt^>(t-s))\xnu(t)\pmu^\leq(s) \nl
\bl+ (\dmu\xnu(t) \dlt^>(s-t))\pnu^\leq(t)\dot\qmu(s)
- \pmu^>(s)\xnu(t) \dds(\dlt^\mu_\nu\dlt^\leq(t-s)) \nl
\bl+ \pnu^>(t)(\dmu\xnu(t)\dlt^>(s-t)) \dot\qmu(s)
- \pmu^>(s) \dds(\dlt^\mu_\nu\dlt^>(t-s))\xnu(t) \\
&&= -\int dt\ \dot\qmu(s)\dmu\xnu(t)\pnu^\leq(t)\dlt(s-t)
 + \dmu\xmu(t)\dlt^>(s-t)\dds\dlt^\leq(t-s) \nl
\bl- \xmu(t)\pmu^\leq(s)\dds\dlt(t-s)
 + \pnu^>(t)\dot\qmu(s)\dmu\xnu(t)\dlt(s-t) \nl
\bl- \dds\dlt^>(t-s)\dmu\xmu(t)\dlt^\leq(s-t)
 -\pmu^>(s)\xmu(t)\dds\dlt(t-s) \nl
&&=	\int dt\ -\no{\dot\qmu(s)\dmu\xnu(t)\pnu(t)} \dlt(s-t)
+\no{\xmu(t)\pmu(s)}\dot\dlt(s-t) \nl
\bl+\dmu\xmu(t) (\dlt^>(s-t)\dot\dlt^\leq(t-s) 
 - \dot\dlt^>(t-s)\dlt^\leq(s-t)) \nl
&&= \fpi (\dmu\ddot\xmu(s) + i\dmu\dot\xmu(s)),
\nnl
&&{[}L^0(s), T(\xi(t),t)] = -\dot\qmu(s)T(\dmu\xi(t),t)\dlt(s-t), 
\label{L0T} \\ \nl
&&{[}L'(s), T(\xi(t),t)] = T(\xi(t),s) \dot\dlt(s-t) \nl 
\bl\pm \fpi T^A_A(\xi(t)) ((2\la-1)\ddot\dlt(s-t) + (2w-1)\dot\dlt(s-t)).
\label{LpT}
\ees
To compute $[L(s), \int dt\ T(\xi(t),t)]$, we note that the
regular pieces from (\ref{L0T}) and (\ref{LpT}) cancel, whereas
the extension acquires the form (\ref{LsLx}). 
The parameters $c'_3$ and $a'_3$ now follows from Lemma \ref{Tlemma}.
\qed
\medskip

The Fock module described in Theorem \ref{Lthm} is reducible,
because it can be decomposed according to the number of $\phi$'s, the
canonical momenta counting negative. If there are several independent
field species, a finer decomposition is possible.
An alternative way to see this is as follows.

Let us refer to $\hat\phim(n)$ and $\hat\pim(n)$ as phase space modes 
of frequency $n$.
The reparametrization generators can be split as 
$L(s) = L^\geq(s)+L^<(s)$, where the {\em raising operators}
$L^\geq(s)$ consist of Fourier modes of non-negative frequency
(as measured by the Hamiltonian (\ref{Ham})),
and the {\em lowering operators} $L^<(s)$ consist of negative ones.
Clearly, every lowering operator contains at least one negative
frequency phase space mode. Because all expressions are normal ordered,
lowering operators thus annihilate the vacuum.
A similar decomposition should be applied to
$\Lxi = \Lxi^\geq + \Lxi^<$, but since $[L(s),\Lxi]=0$ classically,
there are no such lowering operators.

Define a {\em cyclic state} $\ket\emptyset$ to be a state annihilated by
all lowering operators:
\bes
L^<(s)\ket\emptyset = 0.
\ees
As is well known, an irreducible representation	contains only one
cyclic state. Since the vacuum $\ket0$ is cyclic, the existence of
additional cyclic states signals reducibility. The following theorem
describes some cyclic states and their energies, but no claim is made
that the list is exhaustive.

\begin{theorem}
The lowest energy (\ref{le}) of the Fock representation in Theorem 
\ref{Lthm} is
\bes
h = \mp \half \Np \dimm(\rep) ( (w-\half)^2 - (\la-\half)^2 ).
\label{h}
\ees
For a scalar bosonic zero-jet, the state 
$\ket n = (\hat\phi(0))^n\ket0$, $n\geq0$,
is cyclic with energy $h(n) = h+nw$.
For fermionic tensor-valued $p$-jets, set
\bes
\hXi_\ell(n) &=& \prod_\al \prod_{|\mm|\leq\ell} 
 \hat\pham(n), \nl
\hXi_{-\ell-1}(n) &=& \prod_\al \prod_{\ell\leq|\mm|\leq p} 
 \hat\pam(n)
\ees
($\ell\geq0$), where the products run over all components. The states
\bes          
\ket{k,\ell} &=& \hXi_\ell(k-1)\ldots\hXi_\ell(0)\ket0, \nl
\ket{k,-\ell-1} &=& 
 \hXi_{-\ell-1}(k-1)\ldots\hXi_{-\ell-1}(0)\ket0,
\ees
are cyclic, with energy
\bes
h(k,\ell) &=& h + \half {N+\ell\choose\ell} \dimm(\rep)\, (k^2+(2w-1)k), \\
h(k,-\ell-1) &=& h + \half \Big\{ \Np 
 - {N+\ell-1\choose\ell-1}\Big\} \dimm(\rep)\, (k^2-(2w+1)k).
\eens
\end{theorem}

\begin{proof}
Set $\hL(m) = -i\int ds\ e^{ims} L(s)$. 
Then the Virasoro algebra takes the form
\bes
[\hL(m),\hL(n)] &=& (n-m)\hL(m+n) - {c\/12}(m^3-m)\dlt(m+n), \nl
{[}\hL(m),\hat\phim(n)] &=& (n+(1-\la)m+w)\hat\phim(m+n),\\
{[}\hL(m),\hat\pim(n)] &=& (n-\la m-w)\hat\pim(m+n),
\eens
and the Hamiltonian $H=\hL(0)$ (\ref{Ham}).
The action on the vacuum is (excluding the observer)
\bes
\hL'(m)\ket0 = \pm \sum_{n=0}^{m-1} \summp
 (n-\la m+w) \hat\pim(m-n)\hat\phim(n)\ket0.
\label{create}
\ees
To compute parameters, note that
\[
[\hL'(m),\hL'(-m)]\ket0 = (-{c'\/12}m^3 + ({c'\/12}-2h)m) \ket0.
\]
A straightforward calculation shows that $c'$ is given by (\ref{cs})
and $h$ by (\ref{h}).

The property that $\phi(t)$ is a scalar-valued zero-jet is preserved
by $\Lxi$ and $L(s)$. Moreover, any lowering operator gives 
negative-frequency phase space modes when acting on a zero mode, 
and hence the state is cyclic. The energy follows from 
$[\hL(0), \hat\phi(0)] = w\hat\phi(0)$.

Now consider fermions and $\ell>0$. When $\Lxi$ acts on $\hXi_\ell(n)$,
jets of order $|\mm|\leq\ell$ are produced, but no higher-order jets.
Also, $L(s)$ preserves jet order. When acting on $\Xi_\ell(n)$, 
a lowering operator produces a sum of terms, each containing at least 
one phase space mode with frequency
less than $n$, and jet order at most $\ell$. 
However, the state $\ket{k,\ell}$ is the product of all such modes, so
the fermionic property makes all these terms vanish. Hence
$\ket{k,\ell}$ is cyclic. 
The energy $h(k,\ell)$ follows from the following calculation and
Lemma \ref{Tlemma}:
\bes
[\hL(0),\hat\phim(n)] &=& (n+w)\phim(n), \nl
{[}\hL(0),\hXi_\ell(n)] &=& 
 \sum_{|\mm|\leq\ell} (n+w)\hXi_\ell(n) \\
&=&  (n+w) {N+\ell\choose\ell} \dimm(\rep)\, \hXi_\ell(n), \nl
\hL(0)\ket{k,\ell} &=& \big( h + {N+\ell\choose\ell} \dimm(\rep)
 \sum_{n=0}^{k-1} (n+w)\big) \ket{k,\ell}.
\eens
The case $\ell<0$ is completely analogous, except that $\Lxi$ increases
the jet order.  
\qed
\end{proof}

\section{Gauge Algebra}
Consider the gauge algebra $map(N,\oj)$, i.e. maps from
$N$-dimen\-sional spacetime to a finite-dimen\-sional Lie algebra $\oj$,
where $\oj$ has basis $J^a$ (hermitian if $\oj$ is compact and 
semisimple), structure constants $f^{ab}{}_c$, and
Killing metric $\dlt^{ab}$. The brackets are
\bes
[J^a,J^b] = if^{ab}{}_c J^c.
\ees
Let $\dlt^a\propto\tr J^a$ be a priviledged vector satisfying 
$f^{ab}{}_c\dlt^c \equiv 0$. Clearly,
$\dlt^a=0$ if $J^a\in[\oj,\oj]$, but it may be non-zero on abelian 
factors. The primary example is $gl(d)$, where $\tr J^i_j \propto 
\dlt^i_j$.
Our notation is similar to \cite{GO86}.

Let $X=X_a(x)J^a$, $x\in\RR^N$, be a $\oj$-valued
function and define $[X,Y] = if^{ab}{}_c X_aY_bJ^c$.
The generators of $map(N,\oj)$ are denoted by $\J_X$.
The {\em DGRO (Diffeomorphism, Gauge, Repara\-metrization, Observer) 
algebra} $DGRO(N,\oj)$ has brackets
\bes
[\J_X, \J_Y] &=& \J_{[X,Y]} 
 - {1\/2\pi i} (c_5\dlt^{ab} + c_8\dlt^a\dlt^b)
  \int dt\ \dot\qrho(t)\drho X_a(q(t))Y_b(q(t)), \nl
{[}L_f,\J_X] &=& {1 \/4\pi i}\dlt^a 
 \int dt\ (c_6\ddot f(t) - i a_6\dot f(t)) X_a(q(t)),
\label{DGRO} \\
{[}\Lxi, \J_X] &=& \J_{\xmu\dmu X} 
- {c_7 \/2\pi i} \dlt^a\int dt\ \dot\qrho(t) X_a(q(t))\drho\dmu\xmu(q(t)), \nl
{[}\J_X, \qmu(t)] &=& 0, 
\eens
in addition to (\ref{DRO}). 

Alternatively, we can describe $DGRO(N,\oj)$ by the relations
\bes
[\J_X, \J_Y] &=& \J_{[X,Y]} - 
 (c_5\dlt^{ab} + c_8\dlt^a\dlt^b) S_1^\rho(\drho X_aY_b), \nl
{[}L_f,\J_X] &=& \half \dlt^a S_0((c_6 \ddot f - ia_6 \dot f) X_a),
\label{tmap} \\
{[}\Lxi, \J_X] &=& \J_{\xmu\dmu X} 
- c_7\dlt^a S_1^\rho(X_a\drho\dmu\xmu), \nl
{[}\J_X, S_0(F)] &=& [\J_X, S^\rho_1(F_\rho)] = 0,
\eens
in addition to (\ref{tdiff}). 
The cocycle proportional to 
$a_6$ can be removed by the redefinition
\[
\J_X \to \J_X + {ia_6\/2}\dlt^a S_0(X_a),
\]
($\dlt^a[X,Y]_a=0$),
while the remaining terms define non-trivial extensions.
In particular, we recognize the $c_5$ term as the higher-dimensional
generalization of the affine Kac-Moody algebra $\km\oj$.
The present notation has the advantage that all abelian charges
$c_j$, $j = 1, \ldots, 8$, can be discussed collectively.

Let $M$ be a $\oj$ representation. 
We write $T^\mu_\nu = T^\mu_\nu \oplus 1$, $J^a = 1\oplus J^a$,
$1 = 1\oplus1$ for elements in $gl(N)\oplus\oj$, and abbreviate
$M^a = M(1\oplus J^a)$.
$map(N,\oj)$ acts on $J^p\QQ$ and $J^p\PP$
in the following fashion ($V$ indices suppressed):
\bes
[\J_X, \phin(t)] &=& 
 -\summn J^\mm_\nn(X(q(t))) \phim(t),\nl
{[}\J_X, \pim(t)] &=& \summnp 
 \pin(t) J^\mm_\nn(X(q(t))),
\label{Jphi} \\
J^\mm_\nn(X) &\equiv& {\nn\choose\mm} 
 \d_{\nn-\mm} X_a M^a, \nl
{[}\J_X, \qmu(t)] &=& [\J_X,p_\nu(t)] = 0.
\eens
The expression for the matrices  $J^\mm_\nn(X)$, with components 
$J^\mm_\nn(X)^\al_\bt$, follows immediately from
\[
[\J_X, \phin(t)] = \d_\nn( - X_a(q(t)) M^a\phi(q(t))).
\]
They satisfy the following relations:
\bes
J^\mm_{\nn+\mmu}(X) &=& J^\mm_\nn(\dmu X) + J^{\mm-\mmu}_\nn(X), \nl
J^\mm_0(X) &=& \dlt^\mm_0 X_a M^a, \nl 
\dmu J^\mm_\nn(X) &=& J^\mm_\nn(\dmu X), 
\label{Jmn} \\
J^\mm_\nn([X,Y]) &=& \summrn
  J^\rr_\nn(X)J^\mm_\rr(Y) - J^\rr_\nn(Y)J^\mm_\rr(X), \nl
J^\mm_\nn(\xmu\dmu X) &=& \xmu J^\mm_\nn(\dmu X)
 + \summrn T^\rr_\nn(\xi)J^\mm_\rr(X) - J^\rr_\nn(X)T^\mm_\rr(\xi).
\eens
In particular,
$J^\mm_\nn(X) = 0$ if $|\mm|>|\nn|$ and
$J^\mm_\nn(X) = X_a M^a \dlt^\mm_\nn$
if $|\mm|=|\nn|$.

Set $\tr M^a = z_M \dlt^a$ and 
$\tr M^aM^b = y_M \dlt^{ab} + w_M\dlt^a\dlt^b$. For $\oj$ semisimple, 
$w_M = z_M=0$ and $y_M = \psi^2 x_M$, where $\psi$ is the highest
root of $\oj$ and $x_M$ is a positive integer (the Dynkin index of 
the $\oj$ representation $M$) \cite{GO86}.
The analog of Lemma \ref{Tlemma} is
\begin{lemma} \label{Jlemma}
\bes
&i.& \summp \tr J^\mm_\mm(X)
= X_a z_M \dlt^a \Np\dimm(\rep), \nl
&ii.& \sum_{|\mm|,|\nn|\leq p} \tr J^\mm_\nn(X) J^\nn_\mm(Y)
= (y_M\dlt^{ab} + w_M\dlt^a\dlt^b) \Np \dimm(\rep)\, X_aY_b, \nl
&iii.& \sum_{|\mm|,|\nn|\leq p}
 \tr T^\mm_\nn(\xi) J^\nn_\mm(X) \nl
&&= \dmu\xmu X_a z_M \dlt^a (\Np k_0(\rep) + \Npone \dimm(\rep) ).
\eens
\end{lemma}

\begin{proof}
As in Lemma \ref{Tlemma}, only terms with $|\mm|=|\nn|$ contribute to the
sums, and we can hence think of $J^\mm_\nn(X)$ and 
$T^\mm_\nn(\xi)$ as representation matrices in 
$\rep\otimes M\otimes S_\ell$. Hence
\bes
i. &=& \suml \tr M^a\, \dimm(S_\ell) \dimm(\rep),\nl
ii. &=& \suml \tr M^aM^b\, X_aY_b\, \dimm(S_\ell) \dimm(\rep), \nl
iii. &=& \suml X_a \tr M^a\,\dmu\xmu 
 (k_0(\rep)\dimm(S_\ell) + \dimm(\rep)\,k_0(S_\ell)).
\eens
We now apply (\ref{kSp}) and use the definition of $y_M$, $z_M$ and $w_M$. 
\qed
\end{proof}

\begin{theorem}\label{Jthm}
The following operators, together with the operators in 
Theorem \ref{Lthm}, yield a realization of the algebra 
$DGRO(N,\oj)$ on the Fock space $J^p\FF$,
\bes
\J_X &=& \int dt\ J(X(q(t)),t), 
\nlb{JXt}
J(X,t) &=& \mp \summnp  \no{ \pin(t) J^\mm_\nn(X) \phim}(t).
\eens
The parameters are
\bes
c_5 &=& \mp y_M \Np\dimm(\rep), \nl
c_6 &=& \pm z_M \dlt^a (2\la-1) \Np\dimm(\rep), \nl 
a_6 &=& \pm z_M \dlt^a (2w-1) \Np\dimm(\rep), 
\label{Jpar}\\
c_7 &=& \mp z_M \dlt^a (\Np k_0(\rep) + \Npone\dimm(\rep)), \nl
c_8 &=& \mp w_M \Np\dimm(\rep).
\eens
\end{theorem}

\begin{proof}
We use the same notation as in the proof of Theorem \ref{Lthm}.
In particular, capital indices $A=(\al,\mm)$ run over both internal and 
multi-indices, and we write $X(s) = X(q(s))$, etc.
Equation (\ref{JXt}) can be written as
$J(X,s) = J^B_A(X) E^A_B(s)$, where $J^A_B$ satisfy relations (\ref{Jmn})
and $E^A_B(s)$ is as in lemma \ref{FElemma}. 
The following formulas follow immeditately from (\ref{FE}) and (\ref{EE}),
\bes
&&[J(X(s),s),J(Y(t),t)] =
J^B_A(X(s))J^D_C(Y(t)) \times \nl
\bl\times ((\dlt^C_B E^A_D(s) - \dlt^A_D E^C_B(s))\dlt(s-t)
\mp\tpi\dlt^A_D \dlt^C_B\dot\dlt(s-t)) \nl
&&= J([X,Y](s),s)\dlt(s-t) \mp \tpi J^B_A(X(s)) J^A_B(Y(t))\dot\dlt(s-t), 
\nnl
&&[\Lxi^0, J(X(t),t)] = \xmu(t)J(\dmu X(t),t), 
\nnl
&&[T(\xi(s),s),J(X(t),t)] =
T^B_A(\xi(s))J^D_C(X(t)) \times\nl
\bl\times ((\dlt^C_B E^A_D(s) - \dlt^A_D E^C_B(s))\dlt(s-t)
\mp\tpi\dlt^A_D \dlt^C_B\dot\dlt(s-t)) \nle
&&= (J(\xmu(s)\dmu X(t),s) - \xmu(t) J(\dmu X(t),s))\dlt(s-t) \nl
\bl\mp \tpi T^B_A(\xi(s)) J^A_B(X(t)) \dot\dlt(s-t), 
\nnl
&&[L(s),\J_X] \equiv \fpi( g^a \ddot X_a(s) + ib^a\dot X_a(s)), 
\nnl
&&[L^0(s),J(X(t),t)] = -\dot\qmu(s) J(\dmu X(t),t) \dlt(s-t),
\nnl
&&[L'(s),J(X(t),t)] = J^B_A(X(t))( E^A_B(s)\dot\dlt(s-t) \nl
\bl \pm \fpi\dlt^A_B ((2\la-1)\ddot\dlt(s-t) + (2w-1)i\dot\dlt(s-t)) \nl
&&= J(X(t),s) \dot\dlt(s-t) \pm \fpi J^A_A(X(t)) 
((2\la-1)\ddot\dlt(s-t) + (2w-1)i\dot\dlt(s-t)).
\eens
We now collect terms, integrate over $t$, and find that the regular 
terms give the proper algebra,
while Lemma \ref{Jlemma} give the extension parameters.
\qed
\end{proof}

Since $c_5$ must be positive in a unitary represention, the bosonic
Fock space carries a non-unitary representation.

In analogy with (\ref{Lxid}), we can write 
\[
J(X,t) = \mp\summp
 \no{ \pim(t) \dd_\mm(X_a(q(t)) M^a \phi(t))}.
\]
A slight generalization is possible.
The gauge connection corresponds to the jet $A^a_{\mu,\mm}(t)$ 
with conjugate momentum $E_a^{\mu,\mm}(t)$. 
$\tmap(N,\oj)$ acts as
\bes
[\J_X, A^a_{\nu,\nn}(t)] = -\summp
J^\mm_\nn(X(q(t)))^a_b A^b_{\nu,\mm}(t) + \d_{\nn+\nnu} X^a(q(t)),
\label{JA}
\ees
where the matrices $J^\mm_\nn(X)$ are taken in the adjoint representation 
of $\oj$, i.e. $(M^a)^b_c = -if^{ab}{}_c$. 
Thus the contribution to (\ref{JXt}) is
\bes
J(X,t) = \summp \Big\{ \no{ 
E^{\mu,\mm}_b(t) \dd_\mm (i f^{ab}{}_c X_a A^c_{\mu,\mm}(t)) } 
 +  \d_{\mm+\mmu} X^a E^{\mu,\mm}_a(t) \Big\}.
\ees
Due to the non-homogeneous term in (\ref{JA}), the Fock space does not
decompose into subspaces with a fixed number of $A$'s 
as a $\tmap(N,\oj)$ module.

\section{Constraints}
\label{sec:constraints}
Representations of $DRO(N)$ can be restricted to $diff(N)$ using 
techniques from constrained Hamiltonian systems	\cite{Dir64,HT92}.
The same mechanism has appeared in mathematics under
the name Drinfeld-Sokolov reduction \cite{DS85}.

The space $J^p\PP$ is equipped with a natural graded symplectic 
structure, and it can therefore be viewed as a classical phase space.
Let $P,R,...$ label bosonic constraints $\chi_P(q,p,\phi,\pi)$.
If $DRO(N)$ acts in the phase space such that all constraints are 
preserved, we may consider the restriction to the constraint surface
$\chi_P \approx 0$.
Weak equality (i.e. equality modulo constraints) is denoted by $\approx$. 
Constraints are classified as second or first class depending on whether
the Poisson bracket matrix $C_{PR}=[\chi_P,\chi_R]$ is invertible or not. 
First class constraints are connected to gauge symmetries and they 
always generate a Lie algebra.
However, it is often possible to go from
first class to second class (by	fixing a gauge) and back 
(by dropping half the constraints).

Assume that all constraints are second class, if necessary by adding 
gauge-fixing conditions. Then the matrix 
$C_{PR}$ has an inverse, denoted by $\Delta^{PR}$:
\break $\Delta^{PR}C_{RS} = \dlt^P_S$.
The Dirac bracket
\bes
[A,B]^* = [A,B] - [A,\chi_P] \Delta^{PR} [\chi_R,B]
\label{Dirac}
\ees
defines a new Lie bracket which is compatible with the 
constraints: $[A, \chi_R]^* = 0$ for every $A\in C^\infty(J^p\PP)$.

Reparametrizations generate a Lie algebra and can hence be viewed as
first class constraints. A natural gauge choice is to identify
one coordinate with the time parameter. Thus, our constraints are
\[
L(t) \approx 0, \qquad
q^0(t) - t \approx 0.
\]
The Poisson bracket matrix $C(s,t)$ and its inverse $\Delta(s,t)$ are,
on the constraint surface,
\bes
C(s,t) &\equiv& [\chi(s),\chi^T(t)]
= \Big[\begin{pmatrix}q^0(s)-s \\ L(s) \end{pmatrix},
\begin{pmatrix}q^0(t)-t & L(t) \end{pmatrix} \Big] \nl
&\approx& \begin{pmatrix} 0 & \dlt(s-t) \cr -\dlt(s-t) & \virext \end{pmatrix}, \\
\Delta(s,t) &\approx& \begin{pmatrix} \virext & -\dlt(s-t) \cr \dlt(s-t) & 0 \end{pmatrix}.
\eens
We now solve the constraints,
\bes
q^0(t) = t, \qquad
p_0(t) = -q^i(t)p_i(t) + L'(t),
\label{qp0}
\ees
where the latin index $i = 1, 2, ... N-1$ range over 
the remaining (``spatial'') directions.
If $\Lxi$ satisfy the DRO algebra (\ref{DRO}) under the original 
bracket, the Dirac brackets become
\bes
[\Lxi,\Leta]^* &=& \L_{[\xi,\eta]} 
 + \tpi\int dt\  c_1 \dnu\dot\xmu(q(t))\dmu\ynu(q(t)) +\nl
 \bl+ c_2 \dmu\dot\xmu(q(t))\dnu\ynu(q(t)) +\nl
&&+ {1\/4\pi i} \int dt\ 
 c_3 (\dnu\ynu(q(t)) \ddot\xz(q(t))-\dmu\xmu(q(t))\ddot\yz(q(t))) -\nl
\bl -ia_3(\dnu\ynu(q(t)) \dot\xz(q(t))-\dmu\xmu(q(t)) \dot\yz(q(t))) +
\label{gfDRO}\\
&&+ {c_4\/24\pi i} \int dt\ \ddot\xz(q(t))\dot\yz(q(t))
 - \dot\xz(q(t))\yz(q(t)), \nl
{[}\Lxi,\qmu(t)]^* &=& \xmu(q(t)) - \dot\qmu(t)\xz(q(t)), \nl
{[}\qmu(s), \qnu(t)]^* &=& 0, \nl
{[}L(s), \Lxi]^* &=& [L(s),L(t)]^* = [L(s),\qmu(t)]^* = 0,
\eens
where $\dot f(q(t)) = \dot\qrho(t)\drho f(q(t))$ and
\bes
\ddot f(q(t)) = \ddot\qrho(t)\drho f(q(t)) 
 + \dot\qrho(t)\dot\qsi(t)\drho\dsi f(q(t)).
\label{2nd}
\ees
Some other Dirac brackets are
\bes
[\pmu(s), \pnu(t)]^* &=& 
 (\dlt^0_\mu\pnu(s) + \dlt^0_\nu\pmu(t))\dot\dlt(s-t), \nl
{[}\pmu(s), \qnu(t)]^* &=& 
 (\dlt^\mu_\nu - \dot\qmu(t)\dlt^0_\nu) \dlt(s-t), \\
{[}\pmu(s),T(\xi(q(t)),t)]^* &=&
 T(\dmu\xi(q(s)),s) \dlt(s-t) + \dlt^0_\mu T(\xi(q(s)),s) \dot\dlt(s-t).
\eens
Note that $[\Lxi, q^0(t)]^* = 0$. 

Equation (\ref{gfDRO}) is the four-parameter extension of $diff(N)$ found 
in \cite{Lar97a}; it was denoted by $\tdiff(N; c_1,c_2,c_3,c_4)$ in that 
paper. The parameters $c_1$ and $c_2$ are the same 
as in that paper, but I have interchanged the names of the other two:
$c^{\hbox{old}}_3 = 12c^{\hbox{new}}_4$ and 
$c^{\hbox{old}}_4=c^{\hbox{new}}_3$. Note that two of the cocycles are
anisotropic in the sense that they single out the $x^0$ direction.
This anisotropy originates from the gauge choice $q^0(t) \approx t$.
I expect other gauge choices to give rise to even more complicated 
cocycles. Therefore, it is natural to work with the full DRO 
algebra, where the cocycles are of the simple Virasoro form.

Substitution of (\ref{qp0}) into (\ref{emb}) gives
\bes
\Lxi &=& \int dt\ \no{\xi^i(q(t))p_i(t)} - 
\no{\xz(q(t))\dot q^i(t)p_i(t)} 
\nlb{gfix}
&&+ \xz(q(t))L'(t) + T(\xi(q(t)),t), 
\eens
which is the realization found in \cite{Lar97a}. These generators thus
provide an explicit realization of the gauge-fixed algebra (\ref{gfDRO}).
In particular, the Dirac brackets agree with the original brackets
since $q^0(t)$ and $p_0(t)$ have been eliminated.

We can recast (\ref{gfDRO}) as a proper Lie algebra analogous to
(\ref{tdiff}). However, this algebra acquires a very complicated form,
due to the second-order derivatives in (\ref{2nd}).
Not only do the operators
$S_0(F_0)$ and $S^\rho_1(F_\rho)$ enter, but two infinite families of 
linear operators $S_n^\nun(F_\nun)$, $R_n^\rnun(G_\rnun)$,
where $F_\nun(t,x)$, $G_\rnun(t,x)$, $t\in S^1$, $x\in\RR^N$, are
arbitrary functions, totally symmetric in the indices $\nun$.
They have the explicit realization
\bes
S_n^\nun(F_\nun) &=& \tpi \int dt\ \dot q^{\nu_1}(t) .. \dot q^{\nu_n}(t) 
 F_\nun(t,q(t)), \\
R_n^\rnun(G_\rnun) &=& \tpi \int dt\ \ddot \qrho(t) 
 \dot q^{\nu_1}(t) .. \dot q^{\nu_n}(t) G_\rnun(t,q(t)).
\eens
The resulting algebra was written down in \cite{Lar97a}.

The gauge algebra $map(N,\oj)$ is reduced along similar lines.
Since $\J_X$ commutes with both $L(s)$ and $q^0(t)$ 
(before normal ordering), the 
gauge-fixed realization of $map(N,\oj)$ is simply obtained by 
substituting $q^0(t) = t$ in (\ref{JXt}). After normal ordering, the
extension described in \cite{Lar97a} arises, with parameters 
$k = c_5$, $g^a = c_6\dlt^a$ and ${g'}^a = c_7\dlt^a$ given by (\ref{Jpar});
$c_8$ was not considered in that paper.
Realizations of toroidal Lie algebras are obtained by further 
specialization to the $N$-dimensional torus.

\section{Discussion}
The representation theory of diffeomorphism and gauge algebras in more 
than one dimension has been developed. The rather obscure
results in \cite{ERM94,Lar97a} have been given a natural geometric 
explanation in terms of jet space trajectories where the 
reparametrization invariance has been eliminated by gauge fixing. 

These manifestly well defined modules are ``quantum general covariant'',
in the sense that they combine a $diff(N)$ representation 
(general covariance) with the following quantum properties:
Poisson brackets are replaced with commutators, normal ordered expressions
act on a lowest-energy Fock space, and the algebra acquires an extension.
Moreover, these features are obtained without the introduction of
any classical background field. Therefore, these Fock modules can be 
viewed as natural building blocks for theories of quantum gravity.

Classically, everything could be repeated by replacing trajectories by
$d$-dimen\-sional extended objects (e.g. world sheets) in spacetime;
simply reinterpret the variable $t$ in (\ref{emb}) as having $d$ 
components $t^i$. Reparametrization is now expressed by $diff(d)$:
\bes
[L_i(s), L_j(t)] = L_j(s)\d_i\dlt(s-t) 
 + L_i(t) \d_j\dlt(s-t).
\label{diffd}
\ees
However, the quantum theory only exists if $d=1$ (and trivially if $d=0$), 
because otherwise normal ordering
yields infinities and (\ref{diffd}) has no central extension.

\end{document}